\documentclass{Interspeech2024}

\interspeechcameraready 
\usepackage{times}
\usepackage{latexsym}
\usepackage{graphicx}
\usepackage{float}
\usepackage{amssymb}
\usepackage{mathtools}
\usepackage{amsthm}
\usepackage{amsfonts}
\usepackage{subfig}
\usepackage{caption}
\usepackage{subcaption}
\usepackage{amsmath}
\usepackage{algorithm2e}
\usepackage{multirow} 
\usepackage{balance}
\usepackage{cite}

\title{DubWise: Video-Guided Speech Duration Control in Multimodal LLM-based Text-to-Speech for Dubbing}
\name[affiliation={1}]{Neha}{Sahipjohn}
\name[affiliation={1}]{Ashishkumar}{Gudmalwar}
\name[affiliation={1}]{Nirmesh}{Shah}
\name[affiliation={1}]{Pankaj}{Wasnik}
\name[affiliation={2}]{Rajiv Ratn}{Shah}

\address{$^1$Sony Research India Pvt. Ltd., Bangalore, India\\
  $^2$Indraprastha Institute of Information Technology (IIIT), Delhi, India}
  
\email{\{ashish.gudmalwar1,nirmesh.shah,pankaj.wasnik\}@sony.com, rajivratn@iiitd.ac.in}


\begin{document}

\maketitle

\begin{abstract}
Audio-visual alignment after dubbing is a challenging research problem. To this end, we propose a novel method, DubWise: Multi-modal Large Language Model (LLM)-based Text-to-Speech (TTS), which can control the speech duration of synthesized speech in such a way that it aligns well with the speaker’s lip movements given in the reference video even when the spoken text is different or in a different language. To accomplish this, we propose to utilize cross-modal attention techniques in a pre-trained GPT-based TTS. We combine linguistic tokens from text, speaker identity tokens via a voice cloning network, and video tokens via a proposed duration controller network. We demonstrate the effectiveness of our system on the Lip2Wav-Chemistry and LRS2 datasets. Also, the proposed method achieves improved lip sync and naturalness compared to the SOTAs for the same language but different text (i.e., non-parallel) and the different language, different text (i.e., cross-lingual) scenarios.   
\end{abstract}
\section{Introduction}
Recently, AI-based dubbing technologies have become popular with the emergence of LLM-based models like GPT-2 \cite{saeki2021incremental,radford2019language,betker2023better,lajszczak2024base} and XTTS \cite{Eren_Coqui_TTS_2021,xtts}, resulting in high-quality speech synthesis and voice cloning capabilities \cite{caramiaux2019}. The AI-based dubbing technologies first generate the subtitles in the source language with the help of the Automatic Speech Recognition (ASR) service. Then, the Neural Machine Translation (NMT) service converts these subtitles from a source language to the target language. Finally, Text-to-Speech (TTS) technology generates a speech signal in the target language  \cite{yang2020large}. The critical issue with these approaches is that speech generated via TTS in the target language have a different length than the corresponding segment in the source language \cite{wu2023videodubber,chronopoulou2023jointly}. Hence, dubbed audio does not correctly align with the source video content. Hence, the dubbing output look unnatural from an audio-visual synchronization perspective. 

\indent To tackle these alignment issues, one can efficiently utilize time scale modification-based signal processing algorithms, such as, WSOLA and others \cite{verhelst1993overlap,driedger2016review,stylianou2009voice}, to alter the length of synthesized speech. Another possible workaround could be to alter the length of the source video either by downsampling or upsampling video frames via interpolation-based methods \cite{tomar2006converting}. Some approaches have also proposed utilizing phoneme duration predictor networks in TTS to regulate the synthesized speech's overall duration via a multiplication factor \cite{casanova2022yourtts,ren2020fastspeech}. While these methods effectively achieve global duration control, they fail to align audio with reference video. In addition, these methods can alter or control the length of an audio to a certain extent; beyond that, they will significantly deteriorate the quality and intelligibility of speech and, hence, the final dubbing output.


\indent Traditionally, the goal of multimodal TTS systems is to control emotion or speaker-specific information from video modalities \cite{cong2023learning,chen2022v2c,schroeter2000multimodal,liu2021expressive,parker2017expressive}. On the contrary, this paper proposes DubWise, a video-guided speech duration-controllable multimodal TTS for dubbing applications. Our method utilizes visual cues extracted from the video to achieve duration controllability in GPT-based TTS while maintaining intelligibility and speech quality. Videos provide more reliable guidance than audio for alignment in noisy settings. In particular, the proposed architecture utilizes a cross-modal attention technique to combine the video tokens learned via the proposed duration controller network, the audio tokens learned via the voice cloning network, and the linguistic tokens from the input text. The effectiveness of the proposed method has been shown to be superior to the SOTA methods on the Lip2wav Chemistry and LRS2 dataset. 

\indent In summary, our work makes the following contributions:
\begin{itemize}
\item To the best of the authors' knowledge, this is the first attempt of its kind that utilizes video-based modality for achieving duration controllability in autoregressive (AR) Large Language Model (LLM)-based multimodal TTS.
\item Propose a novel duration loss within the auto-regressive model to enable further control over the speech duration.
\item Propose a cross-modal attention technique to integrate video modality against traditional concatenation-based strategies.
\item Propose a lightweight, faster training strategy by only training the randomly initialized cross-attention layers and transpose convolutions layers of the models.
\item Propose to utilize VideoCLIP feature integration to understand global scene context to achieve improved zero-shot voice cloning capability.
\end{itemize}

\section{Related Work}
\textbf{Automatic Video Dubbing (AVD)} 
has been tackled from multiple perspectives.
In \cite{yang2020large}, authors developed the AVD pipeline where, after TTS output, lip movements are modified to match the speech. VideoDubber \cite{wu2023videodubber} approached AVD as a machine translation problem and proposed a translation system based on reference speech duration. Chronopoulou et al. \cite{chronopoulou2023jointly} proposed a model that jointly translates and predicts the durations from speech
SeamlessM4T \cite{barrault2023seamlessm4t} performs speech-to-speech translation while maintaining the prosody of input speech but does not consider alignment. 
Dubbing, alternatively defined as re-recording dialogues due to poor recording conditions, has been addressed by NeuralDubber \cite{hu2021neural}  
Text and the corresponding videos are provided to synthesize speech aligning with the video. Other works like \cite{hassid2022more}, \cite{lu2022visualtts,liu2021expressive} and \cite{cong2023learning} also tackle similar problems. 
Such methods fail when a different text is given in the same or different language. On the contrary, our method works even for scenarios not trained on multilingual videos.
\textbf{Multimodal LLM:} 
Various methods have been suggested for integrating images and videos into text-based LLMs. LLaVA \cite{liu2023visual} suggested that simply by appending a linear layer after a visual representation layer (CLIP) and employing it as a prompt, the vision modality can be integrated into LLMs. On the other hand, Flamingo \cite{alayrac2022flamingo} presents a visual language model where the vision modality is incorporated into the prompt, and cross-attention layers within the LLM are introduced and trained accordingly. Other approaches focus on image and video comprehension and storytelling. 
However, our method introduces controllability to GPT-based TTS systems by leveraging video input. GPT based TTS primarily utilizes LLM based GPT decoder and we have used both LLM and GPT term interchangeably throughout this paper.
\section{Proposed Method}
\begin{figure*}
  \centering
  \includegraphics[width=0.75\linewidth]{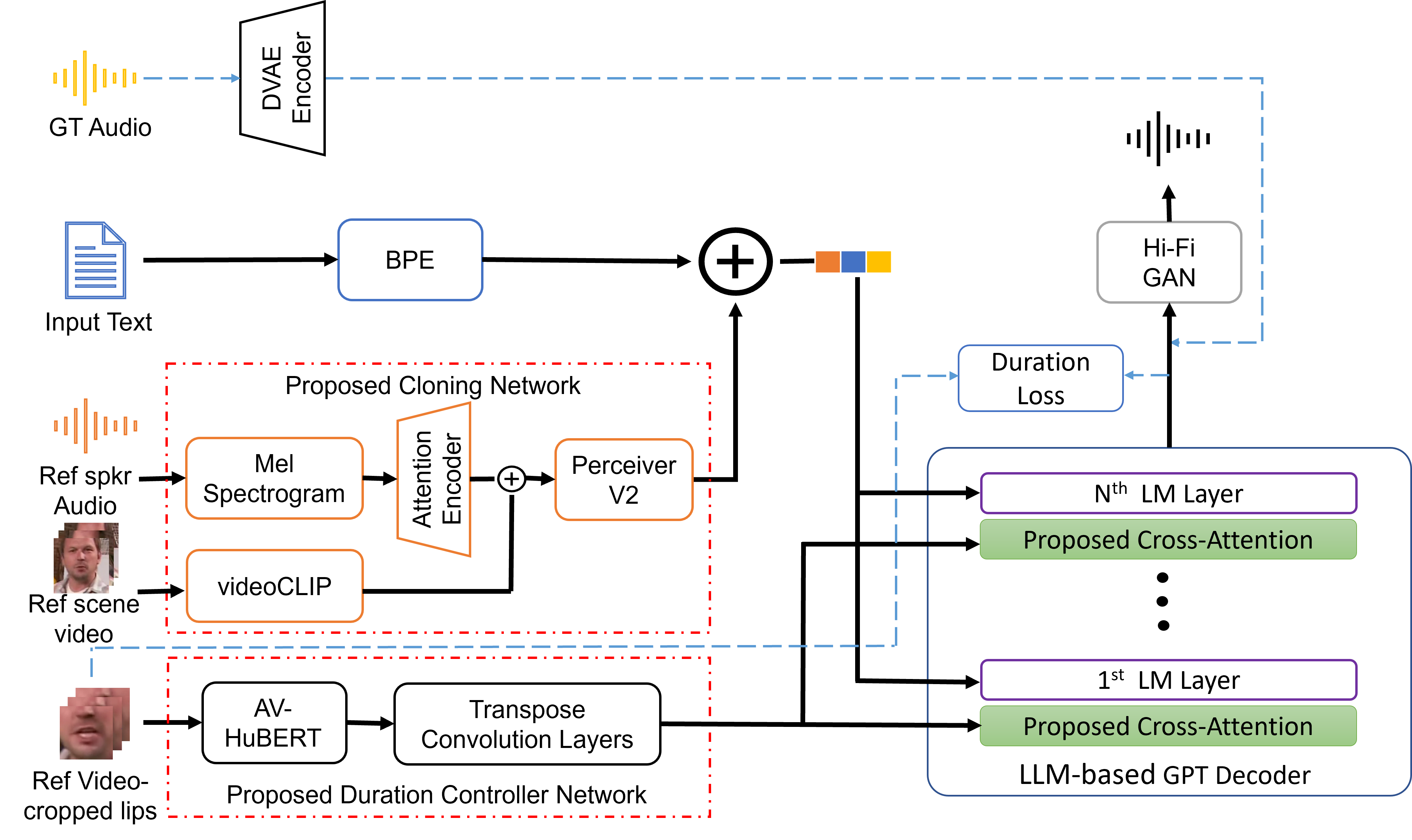}
  \vspace{-0.3cm}
  \caption{Proposed Method: Tokenized reference-speaker audio and text form the model's prompt (ground truth audio included during training only). Lip region video is fed through cross-attention. HiFi-GAN generates speech from the output.}
  \vspace{-0.5cm}
  \label{fig:block_diagram}
\end{figure*}
The core of our model is a pre-trained, multilingual TTS based on GPT2 architecture. The key reasons for preferring GPT2 over its recent architectures are smaller model size and its wider adaptability in the SOTA TTS architectures \cite{radford2019language,betker2023better,lajszczak2024base}. 
We integrate lip-reading features to direct the TTS output in Figure \ref{fig:block_diagram}, aiming to attain synchronization between the generated speech and the observed lip movements in the reference video. 
DubWise models the speech $S$ conditioned on lip-features $F_{lip}$ and input text $C_t$.
\vspace{-0.3cm}  
\begin{equation}
    p(S | C_t, F_{lip}) = \prod_{l=1}^{L} p(S_{l} | S _{<l}, C_t, F_{lip})
\end{equation}
 where $ C_t = translate(C_s)$ is translated text for source text ($C_s$).

We use an LLM-based TTS, XTTS \cite{Eren_Coqui_TTS_2021} as the backbone. The system utilizes byte-pair tokenizer to encode input text.  
Reference speaker audio $S_{r}$ uttering arbitrary sentence is used to clone the voice.
We employ a speaker encoder network to capture the speaker's unique traits. This network extracts speaker embeddings, which a perceiver \cite{jaegle2021perceiver} subsequently processes to derive a fixed-dimensional representation. 
To capture the overall scene and mood of the utterance, videoCLIP \cite{xu2021videoclip} features of a reference video is used. This is given to the perceiver along with the speaker embedding to capture overall speaker style, while also training the perceiver.
A one-hot vector-based language ID is used to specify the language of the input text.
These three elements – language ID $L_t$, speaker embedding $S_{r}$, and the input text  $C_t$ – are then concatenated to form the prompt, which is subsequently fed into the GPT model. 
This entire system is pre-trained on 16 languages. 
During training, the GPT model learns to predict the next token autoregressively. A pre-trained DVAE  is utilized to tokenize the audio data for GPT model training.

\textbf{Video-encoder:} 
Given a reference video of a person speaking in a source language $L_s$ that you want to dub to a target language $L_t$, the cropped lip-region $V_{lip}$ from the silent source video is retrieved. The lip-representation features $F_{lip}$ are extracted using lip-reading model $M_{lip}$.
This work employs a lip encoder model trained on masked prediction-based self-supervised learning. We leverage the SOTA pre-trained AV-HuBERT \cite{shi2022learning} model as the lip feature extractor. 
During training, this model takes both video (cropped lip region) and audio as inputs. 
 A selection of audio and video frames are masked, and the transformer component within the model is trained to predict the iteratively refined cluster IDs. The initial cluster IDs are derived through k-means clustering of Mel-Frequency Cepstral Coefficients (MFCC) features extracted from the audio. 
Finally, the model undergoes fine-tuning with only video features as input, using a transformer decoder designed for lip-reading task to predict text from observed lip movements. 

We utilize these features extracted by the AV-HuBERT model before the Language Model (LM) Transformer head as the lip features. 
\begin{equation}
    F_{lip} = M_{lip}(V_{lip}) 
\end{equation}


\textbf{Integration:} 
This work explores two approaches to incorporating video features into a TTS system for introducing video-based controllability. The first approach involves concatenating video features with speaker embeddings and text into a single prompt fed to the GPT model. This results in an extended prompt length, and the entire GPT model undergoes fine-tuning based on this modified prompt structure.
\begin{equation}
    prompt = [L_t, S_{ref}, F_{lip}, C_t]
    \vspace{-0.2cm}
\end{equation}
The second approach leverages cross-attention mechanisms within the LLM architecture.
We introduced cross-attention layers between the GPT model, which directly attend to the video features.
To ensure the generated speech duration aligns with the reference video, we extend the video features to match the expected speech duration.
This simplifies training by enabling linear attention towards the video features without requiring explicit duration specification. We achieve this by applying a combination of convolution and transposed convolution to the video features, effectively learning the necessary expansion without manual intervention. An end token is then concatenated to mark the sequence termination. During training, all other model parameters remain fixed, while only the cross-attention layer and the newly introduced transposed convolution layers are fine-tuned for the language modeling task.

\par The losses in the model include cross-entropy loss on audio-tokens ($CE_{audio}$), scaled ($\alpha$) text tokens loss ($CE_{text}$) and scaled ($\beta$) duration loss ($duration\_loss$). As the embedding layers remain pretrained and unaltered, they can be used to infer predicted token IDs. We identify the end of a sequence by detecting the occurrence of a pre-defined stop token within the predicted sequence and ground-truth audio sequence. The proposed duration loss  is calculated as the difference between the index of the end token in ground-truth audio and the predicted end token.  
\begin{equation}
    loss = CE_{audio} + \alpha * CE_{text} + \beta * duration\_loss \\
\end{equation}
The HiFi-GAN \cite{kong2020hifi} vocoder-based decoder was utilized to extract audio from the GPT output's latent representation.
\section{Experiments}
This section provides details of the datasets and evaluation metrics employed for evaluation. 

\vspace{0.2cm}
\indent \textbf{Datasets:}
The experiments are performed on Lip2Wav-chemistry dataset \cite{hu2021neural},\cite{prajwal2020learning}. It is a single-speaker, 9.2 hours long, YouTube lecture video  based dataset. We also perform experiments on multi-speaker LRS2 dataset \cite{Afouras18c}, which is a 29 hours long YouTube clips dataset. To extract the lip-reading features, the video sequences are resampled to a frame rate of 25 frames per second (fps). We utilize the $S^3FD$ \cite{zhang2017s3fd} face detector to detect facial key points, and crop a mouth-centered region-of-interest (ROI). This is used to extract the lip-reading features. The corresponding text is tokenized using Byte-pair-encoding. 
For evaluations using translated text, we generate the translation using IndicTrans2 \cite{gala2023indictrans}, an off-the-shelf machine translation model.

\indent \textbf{Baseline Methods:} We compare our results with following SOTA methods.
    FastSpeech2  \cite{ren2020fastspeech}
    proposed phoneme based global duration controllability in  the generated audio by speed up or slowed down by a factor. 
    YourTTS \cite{casanova2022yourtts} leverages a stochastic duration predictor while training.
    XTTS \cite{Eren_Coqui_TTS_2021} is GPT2-based TTS model which doesn't explicitely have duration controlability.
    To adjust the length of XTTS-generated speech, we employ the Wave Similarity Overlap and Add (WSOLA) algorithm \cite{verhelst1993overlap} within the FFmpeg toolkit. This approach is referred to as $XTTS$+$WSOLA$.
    HPMDubbing \cite{cong2023learning}  is a multimodal dubbing pipeline for same-text scenarios, where corresponding text and video are inputs. We evaluate it for same-text and different-text scenarios.
    
 
 

\indent \textbf{Evaluation Metrics:}
We do both subjective and objective evaluations to demonstrate the efficacy of the model. In objective evaluation, duration modelling capability, is measured using the ratio of durations of synthesized speech to reference video (DR, ideal value is 1) and difference between durations of synthesized speech  and reference video (DD, ideal value is 0)
To evaluate the intelligibility  and performance of TTS, we calculate the word error rate (WER) and character error rate (CER).
Alignment of  speech to lip-movements - lip-sync - is evaluated using  Lip Sync Error - Confidence (LSE-C) and Lip Sync Error - Distance (LSE-D) \cite{prajwal2020lip}. AV-Offset is the time offset between audio and video.  The measurements are taken using a pretrained
SyncNet model \cite{chung2017out}.
In subjective evaluation, we compute the Mean Opinion Score (MOS) on overall quality and intelligibility for the speech samples generated by different methods. 20 subjects (aged between 25 to 35 years and no known hearing impairment) took part in the subjective evaluation.
\section{Results}
We present results in three different scenarios, namely, same-text, different-text, and cross-lingual, as shown in Table \ref{tab:tab1}. Demo samples are available at \footnote{\url{https://nirmesh-sony.github.io/DubWise/}}. In the same-text condition, 
the same text as in the reference video is input, to synthesize audio samples. On the other hand, in different-text condition, audio samples are synthesized using text different from that in the videos. Here, we changed the sentence structure by $\pm$ 50\% in terms of word-count from the video's original text. We also evaluated the proposed approach in a cross-lingual setting in which Hindi audio is generated by using an English speaker as a reference for voice cloning.

For the same text condition, Dubwise achieves the best DR and DD close to ground truth except for FastSpeech2 and XTTS$+$WSOLA. We can see that overall global duration controllability is achieved by speeding up or slowing down the FastSpeech2 and XTTS$+$WSOLA TTS output. However, it introduces distortion and affects the intelligibility of speech, which is evident from the higher WER and CER scores compared to DubWise. Regarding intelligibility evaluation, Dubwise outperforms other baselines in terms of WER and CER. On the LSE evaluation side, Dubwise achieves the best LSE-D and LSE-C scores except for the HPMDubbing. HPMDubbing is trained to synthesize audio based on the audio-video pair using the same text in the video. It tries to lip-sync with audio for the same text, but it fails to maintain intelligibility and constrains the duration of generated audio, which is evident from the high WER, CER, DR, and DD. The proposed DubWise overcomes this by maintaining intelligibility and speech quality while minimizing duration differences. The proposed approach also tries to achieve lip sync at the same time. 


\indent Different-text scenario (Table \ref{tab:tab1}), also displays similar trebd as above. 
Our method has a lower WER and CER with a decent duration ratio. 
In terms of Lip-Sync-Error, DubWise achieves the best score for LSE-D and LSE-C, apart from the HPMDubbing approach. HPMDubbing's WER and CER scores are way higher as they are highly unintelligible and distorted. This is due to HPMDubbing being trained to synthesize audio using audio-video pairing with the same text as in the video. Another drawback of HPDubbing is it does not have cross-lingual support.

\indent Similar trend is found for cross-lingual scenario, where audio is synthesized using Hindi text and English reference video to control the generated audio duration. The DubWise approach preserves intelligibility and speech quality while minimizing variations in duration. Additionally, the proposed approach aims to achieve lip synchronization simultaneously. As HPMdubbing and FastSpeech2 doesnot support cross-lingual support, we could not compare DubWise approach with them. 
\begin{table}[!h]
\vspace{-0.2cm}
\caption{Objective evaluation on Lip2Wav-Chem dataset}
\vspace{-0.3cm}
\label{tab:tab1}
\resizebox{1\linewidth}{!}{
\begin{tabular}{cccccccc}   \hline
\multirow{2}{*}{Method}                                                  & \multicolumn{2}{c}{Duration Control}                                                                                                 & \multicolumn{2}{c}{Intelligibility}                           & \multicolumn{3}{c}{Lip Sync Error}                                                                    \\
    & \begin{tabular}[c]{@{}c@{}}DR
    \end{tabular} & \begin{tabular}[c]{@{}c@{}}DD
    \end{tabular} & WER $\downarrow$ & CER $\downarrow$ & LSE-D $\downarrow$ & LSE-C $\uparrow$ & AV offset $\downarrow$  \\  \hline

\multicolumn{8}{c}{Same Text}
\\ \hline

Ground Truth                                                             & 1                                                                  & 0                                                                       & 4.05                          & 2.02                          & 7.32                            & 7.19                          & 0.6                              \\
FastSpeech2                                                              & 0.94                                                               & 0.29                                                                    & 7.87                          & 2.79                          & 12.76                           & 2.37                          & 7.43                              \\
\begin{tabular}[c]{@{}c@{}}YourTTS\\ \end{tabular}     & 1.12                                                               & 0.53                                                                    & 27.0                          & 14.85                         & 13.01                           & 1.48                          & 7.27                           \\
Baseline\_XTTS2                                                          & 1.26                                                               & 1.335                                                                   & 8.1                           & 4.68                          & 13.01                           & 2.25                          & 8.15                         \\
XTTS2+WSOLA                                                              & \textbf{0.99}                                                              & \textbf{0.019}                                                                   & 9.19                          & 4.38                          & 12.96                           & 2.12                          & 8.16                             \\
HPMDubbing & 0.837&0.724 &6.29&3.75&\textbf{6.97}& \textbf{7.17}& \textbf{3.97}  \\
\begin{tabular}[c]{@{}c@{}}DubWise (Ours)\\ \end{tabular}                 & 1.077                                                            & 0.432                                                                   & \textbf{4.13}                          & \textbf{2.48}                          & 12.1                        & 2.82                  &      5.98  \\  \hline           
\multicolumn{8}{c}{Different Text}
\\ \hline
FastSpeech2                                                              & 0.947                                                              & 0.23                                                                    & 22.48                         & 13.28                         & 13.65                           & 1.61                          & 9.53                                \\
\begin{tabular}[c]{@{}c@{}}YourTTS\\ \end{tabular}     & 1.014                                                              & 0.597                                                                   & 31.92                         & 18.49                         & 13.36                           & 1.61                          & 8.7                                 \\
XTTS                                                        & 1.29                                                               & 1.92                                                                    & 16.72                         & 10.19                         & 13.51                           & 1.78                          & 9.67                               
   \\
XTTS$+$WSOLA                                                              & \textbf{0.995}                                                              &\textbf{0.019}                                                                  & 21.17                         & 13.08                         & 13.39                           & 1.64                          & 9.08                               \\

HPMDubbing&0.784&1.026&119.01&92.82&\textbf{8.94}&\textbf{6.40}&\textbf{4.18}     \\
\begin{tabular}[c]{@{}c@{}}DubWise (Ours)\\ \end{tabular}                 & 0.94                                                               & 1.03                                                                    & \textbf{11.94}                         & \textbf{8.13}                          & 13.31                        & 1.91                        & 9.54     \\   \hline  

\multicolumn{8}{c}{Different Language Text}
\\ \hline
                         \\
\begin{tabular}[c]{@{}c@{}}YourTTS\\ \end{tabular} & 1.12                                                               & 0.69                                                                    & 56.23                         & 43.85                         & 13.43                           & 1.55                          & 8.21                             \\
XTTS                                                     & 1.32                                                               & 1.46                                                                    & 34.34                         & 25.17                         & 13.44                           & 1.67                          & 10.14                            \\
XTTS$+$WSOLA                                                          & \textbf{0.99}                     & \textbf{0.018}                                                                   & 38.38                         & 29.06                         & 13.39                           & 1.65                          & 9.35                             \\
\begin{tabular}[c]{@{}c@{}}DubWise (Ours)\\ \end{tabular}             & 1.19            & 0.932                                                                   & \textbf{22.76}                         & \textbf{13.59}                         & \textbf{13.33}                           & \textbf{1.58}                          & \textbf{8.91}     \\  \hline                               
\end{tabular}}
\end{table}
\begin{table}[ht!]
\vspace{-0.3cm}
\caption{Objective evaluation on LRS2 dataset}
\vspace{-0.3cm}
\label{tab:lrs2table}
\resizebox{1\linewidth}{!}{
\begin{tabular}{cccccccc}
\hline

 \multirow{2}{*}{Method} & \multicolumn{2}{c}{Duration Control} & \multicolumn{2}{c}{Intelligibility}  & \multicolumn{3}{c}{Lip Sync Error} \\
   & \begin{tabular}[c]{@{}c@{}}DR
   \end{tabular} & \begin{tabular}[c]{@{}c@{}}DD
   \end{tabular} & WER & CER  & LSE-D $\downarrow$ & LSE-C $\uparrow$ & AV offset $\downarrow$ \\ \hline

\multicolumn{8}{c}{Same Text}
\\ \hline

Ground truth & 1 & 0 & 33.04&27.16&6.23&8.41 &0.06  \\
 XTTS           & 1.96 & 1.41 & \textbf{22.10} & \textbf{15.45} & 11.55&3.02&8.27  \\
 DubWise (Ours) &\textbf{ 1.18} & \textbf{0.28} & 38.88&29.35& \textbf{10.16}&\textbf{4.35}&\textbf{3.56} \\ \hline

\multicolumn{8}{c}{Different Language Text}
\\ \hline
 
  XTTS & 3.15 & 3.13 & 49.24 & 40.22 &12.23&2.38&10.33  \\
 DubWise (Ours) & \textbf{1.29} &\textbf{ 0.46} & \textbf{46.77} & \textbf{36.53} & \textbf{11.60}&\textbf{2.86}&\textbf{7.61}
 \\ \hline
\end{tabular}}
\vspace{-0.55cm}
\end{table}
\begin{figure}[h!]
  \centering
  \includegraphics[width=0.9\linewidth]{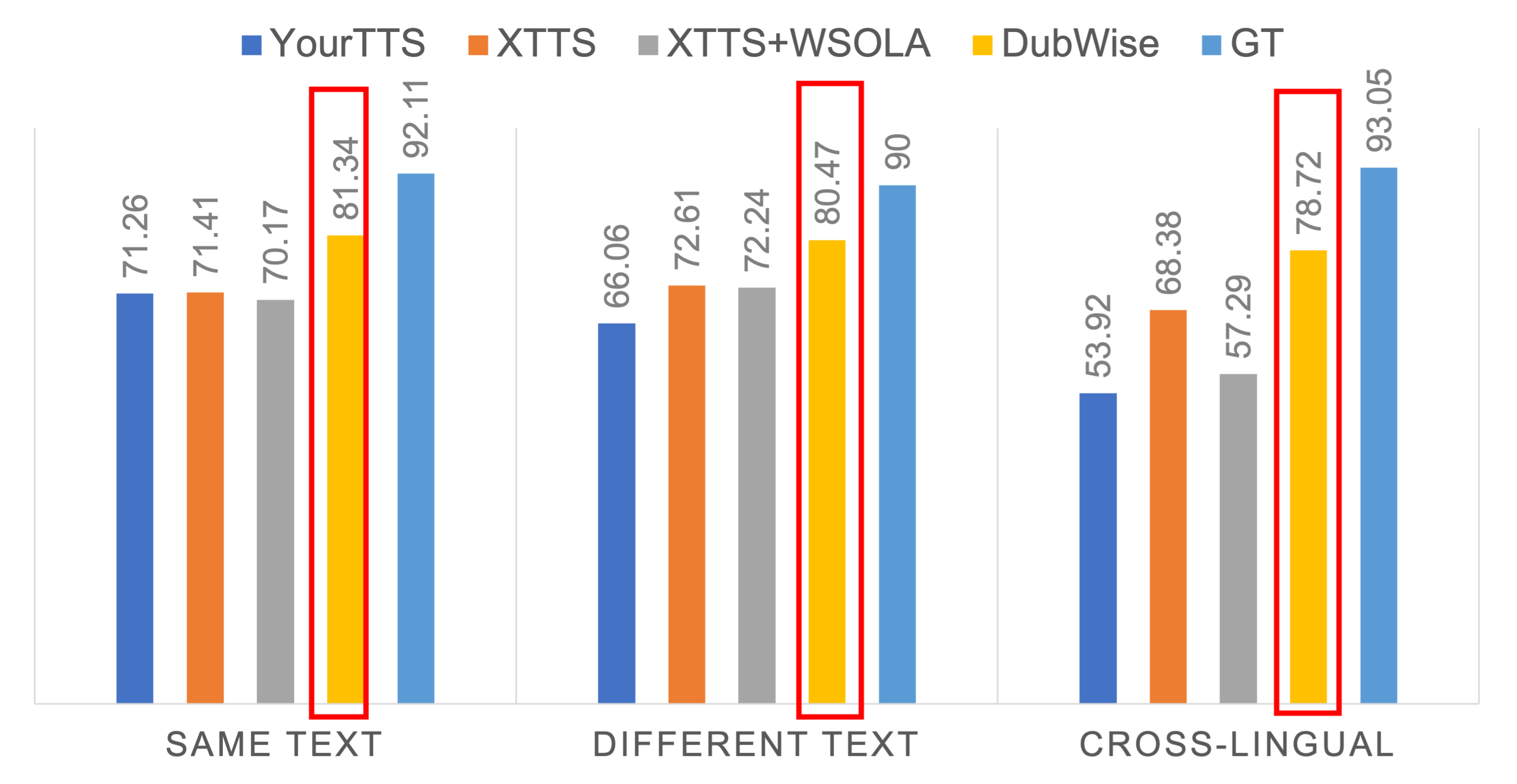}
  \vspace{-0.3cm}
  \caption{The subjective evaluation of proposed DubWise approach and baseline methods. We obtained $p$-value $<$ 0.004}
  \vspace{-0.3cm}
  \label{fig:subjective_eval}
\end{figure}

To demonstrate multispeaker capabilities, we also evaluate the LRS2 dataset for same-text and translated Hindi text scenarios and compare the duration controllability aspect with respect to the baseline XTTS model. We observe that DubWise is able to achieve good values for duration difference and duration ratio. The WER is high due to noisy data and small samples used for speaker-reference voice. We also see that DubWise achieves almost 100\% $SLC_p$ in both same-text and translated-text scenarios. From subjective evaluation, we can see that it also supports observations in the objective evaluation. In particular, the proposed DubWise approach achieves the highest score for audio quality, as shown in Figure \ref{fig:subjective_eval}. 

\vspace{-0.2cm}
\subsection{Ablation Experiments}
We investigated different architectures for the DubWise model through ablation analysis (refer to Table \ref{tab:ablation}). 
In Ablation 1, we integrated the video features by concatenating them to the input prompt of GPT-TTS. But this resulted in a higher Word Error Rate (WER) compared to other methods.  Ablation 2  using cross-attention to integrate video information with GPT-TTS, achieved the lowest WER and lip-sync error.
In Ablation 3, videoCLIP features were incorporated for global style control, but this led to performance degradation compared to Ablation 2.
Ablation 4 incorporated duration loss with cross-attention, which improved Duration Ratio compared to Ablation2.  
By incorporating duration loss and videoCLIP with cross-attention, DubWise achieved the best balance between speech quality (measured by WER and lip-sync error) and control over speech duration (DR). The percentage of sentences matching the desired duration was found to be maximum in this case.
\begin{table}[ht!]
\caption{Ablation Study}
\vspace{-0.3cm}
\label{tab:ablation}
\scalebox{0.6}{
\begin{tabular}{cccccccc}   \hline
\multirow{2}{*}{Method} & \multicolumn{2}{c}{Duration Control}                                                                    & \multicolumn{2}{c}{Intelligibility}                           & \multicolumn{3}{c}{Lip Sync Error}                                                               \\
                        & \begin{tabular}[c]{@{}c@{}}DR\\ (Ideal$=$1)\end{tabular} & \begin{tabular}[c]{@{}c@{}}DD\\ (Ideal$=$0)\end{tabular} & WER $\downarrow$ & CER $\downarrow$ & LSE-D $\downarrow$ & LSE-C $\uparrow$ & AV offset $\downarrow$   \\    \hline
Ablation1               & 1.07                                                   & \textbf{0.43}                                          & 8.94                          & 5.92                          & 11.93                           & 3.01                          & 5.71                            \\
Ablation2               & 1.06                                                   & \textbf{0.43}                                                  & \textbf{4.39}                 & \textbf{2.20}                 & \textbf{11.84}                 & \textbf{3.07}                 & \textbf{5.52}              \\
Ablation3               & 1.08                                                   & 0.46                                                   & 4.63                          & 2.50                          & 12.11                           & 2.79                          & 5.71                              \\
Ablation4               & 1.06                                                   & 0.45                                                   & 4.63                          & 2.50                          & 12.15                           & 2.83                          & 5.69                              \\
DubWise                 & \textbf{1.04}                                          & 0.45                                                 & 4.86                          & 2.73                          & 12.07                           & 2.84                          & 5.77                \\   \hline             
\end{tabular}}
\vspace{-0.25cm}
\end{table}

\begin{figure}[h!]
  \centering
  \includegraphics[height=4.5cm, width=\linewidth]{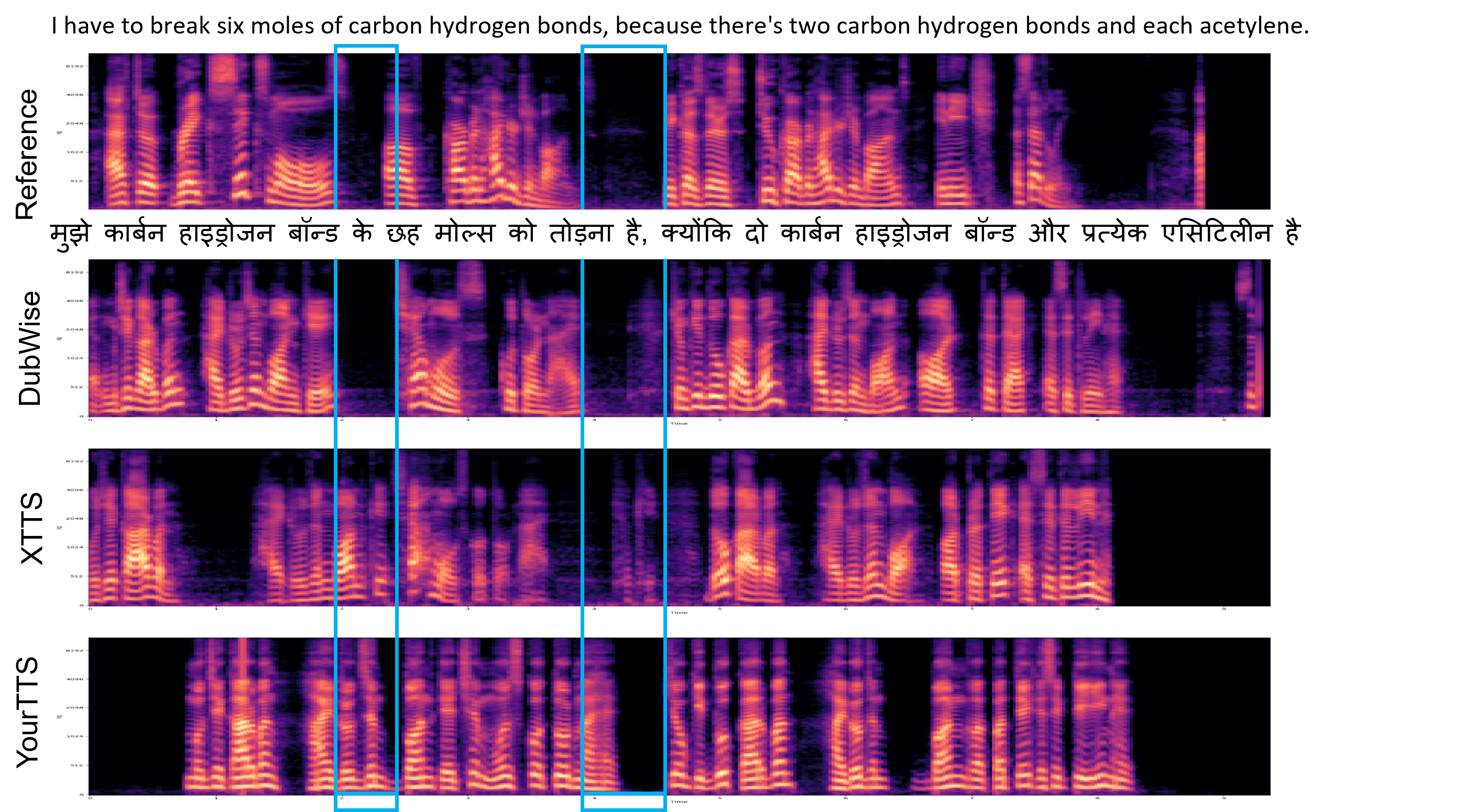}
  \vspace{-0.5cm}
  \caption{Spectrographic analysis of the reference English speech, and corresponding translated Hindi synthesized speech using DubWise and different baselines. Here, reference English sentence is ``I have to break six moles of carbon hydrogen bonds, because there's two carbon hydrogen bonds and each acetylene.''} 
  \vspace{-0.3cm}
  \label{fig:mel}
\end{figure}

\indent Figure \ref{fig:mel} illustrates an example of a mel-spectrogram, demonstrating how the proposed model captures pauses (highlighted within the blue box) and lip movements in the synthesized speech in the considered cross-lingual scenario compared to the SOTA algorithm. 
\vspace{-0.3cm}
\section{Summary and Conclusion}
In this paper, we propose a novel method, DubWise: Multi-modal Large Language Model (LLM)-based Text-to-Speech (TTS), which can control the speech duration of synthesized speech in such a way that it aligns well with the speaker’s lip movements given in the reference video even when the spoken text is different  or in a different language. To accomplish this, we proposed to utilize cross-modal attention techniques in a pre-trained GPT-based TTS, while combining linguistic tokens from text, speaker identity tokens via voice cloning network and video tokens via proposed duration controller network. We demonstrate the effectiveness of our system on Lip2Wav-Chemistry and LRS2 datasets. Also, the proposed method achieves improved lip sync and naturalness compared to the SOTAs for the same language but different text (i.e., non-parallel) and the different language different text (i.e., cross-lingual) scenarios. In future, one should work on improving the global duration controllability as well as word level and phrase-level duration controllability. 
\bibliographystyle{IEEEtran}
\balance
\bibliography{mybib}

\end{document}